\documentclass[aps,prx,twocolumn,amsmath,amssymb,nofootinbib,superscriptaddress]{revtex4-2}
\usepackage{times}
\usepackage{braket}
\usepackage[pdftex]{graphicx}
\usepackage{dcolumn}
\usepackage{bm}
\usepackage{bbm}
\usepackage{amsmath}
\usepackage{indentfirst}
\usepackage[colorlinks,linkcolor=blue,anchorcolor=blue,citecolor=blue]{hyperref}
\usepackage[dvipsnames]{xcolor}
\usepackage{xcolor}

\colorlet{shadecolor}{gray!40}

\usepackage{subfigure}
\usepackage{algorithm}
\usepackage{algorithmicx}
\usepackage{algpseudocode}
\usepackage{amsmath}
\usepackage{verbatim}
\usepackage{makecell}
\usepackage[normalem]{ulem}
\usepackage{setspace}
\usepackage{float}
\usepackage{booktabs}
\usepackage{threeparttable}

\begin{document}
\title{Scalable Constant-Time Logical Gates for Large-Scale Quantum Computation Using Window-Based Correlated Decoding}

\author{Jiaxuan Zhang}
\affiliation{Key Laboratory of Quantum Information, Chinese Academy of Sciences, School of Physics, University of Science and Technology of China, Hefei, Anhui, 230026, P. R. China}

\author{Zhao-Yun Chen}
\affiliation{Institute of Artificial Intelligence, Hefei Comprehensive National Science Center, Hefei, Anhui, 230088, P. R. 
 China}

\author{Jia-Ning Li}
\affiliation{Key Laboratory of Quantum Information, Chinese Academy of Sciences, School of Physics, University of Science and Technology of China, Hefei, Anhui, 230026, P. R. China}

\author{Tian-Hao Wei}
\affiliation{Key Laboratory of Quantum Information, Chinese Academy of Sciences, School of Physics, University of Science and Technology of China, Hefei, Anhui, 230026, P. R. China}

\author{Huan-Yu Liu}
\affiliation{Key Laboratory of Quantum Information, Chinese Academy of Sciences, School of Physics, University of Science and Technology of China, Hefei, Anhui, 230026, P. R. China}

\author{Xi-Ning Zhuang}
\affiliation{Key Laboratory of Quantum Information, Chinese Academy of Sciences, School of Physics, University of Science and Technology of China, Hefei, Anhui, 230026, P. R. China}

\author{Qing-Song Li}
\affiliation{Key Laboratory of Quantum Information, Chinese Academy of Sciences, School of Physics, University of Science and Technology of China, Hefei, Anhui, 230026, P. R. China}

\author{Yu-Chun Wu}
\email{wuyuchun@ustc.edu.cn}
\affiliation{Key Laboratory of Quantum Information, Chinese Academy of Sciences, School of Physics, University of Science and Technology of China, Hefei, Anhui, 230026, P. R. China}
\affiliation{Institute of Artificial Intelligence, Hefei Comprehensive National Science Center, Hefei, Anhui, 230088, P. R. China}

\author{Guo-Ping Guo}
\email{gpguo@ustc.edu.cn}
\affiliation{Key Laboratory of Quantum Information, Chinese Academy of Sciences, School of Physics, University of Science and Technology of China, Hefei, Anhui, 230026, P. R. China}
\affiliation{Institute of Artificial Intelligence, Hefei Comprehensive National Science Center, Hefei, Anhui, 230088, P. R. China}
\affiliation{Origin Quantum Computing Hefei, Anhui 230026, P. R. China}
\date{\today}
\begin{abstract}
Large-scale quantum computation requires to be performed in the fault-tolerant manner. One crucial challenge of fault-tolerant quantum computing (FTQC) is reducing the overhead of implementing logical gates. Recently work proposed correlated decoding and ``algorithmic fault tolerance" to achieve constant-time logical gates that enables universal quantum computation. However, for circuits involving mid-circuit measurements and feedback, the previous scheme for constant-time logical gates is incompatible with window-based decoding, which is a scalable approach for handling large-scale circuits. In this work, we propose an architecture that employs delayed fixup circuits and window-based correlated decoding, realizing scalable constant-time logical gates. This design significantly reduces both the frequency and duration of decoding, while maintaining support for constant-time and universal logical gates across a broad class of quantum codes. More importantly, by spatial parallelism of windows, this architecture well adapts to time-optimal FTQC, making it particularly useful for large-scale quantum computation. Using Shor's algorithm as an example, we explore the application of our architecture and reveals the promising potential of using constant-time logical gates to perform large-scale quantum computation with acceptable overhead on physical systems like ion traps.
\end{abstract}

\maketitle
\section{Introduction}
Quantum computing offers the potential to tackle complex problems such as large integer factorization~\cite{365700} and quantum simulation~\cite{Feynman+1988+523+548,freedman2002simulation}. Large-scale quantum computation relies on quantum error correction (QEC)~\cite{preskill1998reliable,gottesman1997stabilizer} and fault-tolerant quantum computing (FTQC)~\cite{548464,campbell2017roads}. Fault-tolerant gate operations on logical qubits are typically much more complex than operations on physical qubits, consuming substantial space-time overhead. Therefore, developing low-overhead logical gates is a crucial challenge in driving FTQC toward practical implementation. In the conventional FTQC framework, error correction follows each logical gate to prevent error accumulation and propagation~\cite{nielsen2010quantum}, typically involving $\Theta(d)$ rounds of syndrome extraction, where $d$ is the code distance. Consequently, the time of a logical gate implementation is directly relevant to $d$.

Based on correlated decoding, Ref.~\cite{PhysRevLett.133.240602} indicated that only one round of syndrome extraction is required after transversal gates, thereby achieving constant-time transversal gate. The fast implementation of transversal gates reduces the overall space-time overhead in fault-tolerant quantum computing (FTQC). However, transversal gates of any QEC code are not universal~\cite{physrevlett.102.110502,physreva.78.012353,6006592}. For example, for the surface code, a transversal gate set includes $\{H, S, \text{CNOT}\}$~\cite{PhysRevA.94.042316}, assuming a platform with all-to-all connectivity, such as trapped ions~\cite{da2024demonstration,physrevx.11.041058} or neutral atoms~\cite{bluvstein2024logical}. While the implementation of logical $T$ gates requires magic states and subsequent fixup operations $S$ according to the mid-circuit measurement outcomes~\cite{PhysRevA.71.022316}. Directly applying constant-time transversal gates in circuits involving mid-circuit measurement and feedback is infeasible, since there are insufficient rounds of syndrome extraction to ensure reliable decoding results before making feedback decisions. The decoding result immediately after mid-circuit measurements is possibly inconsistent with subsequent decoding results obtained after sufficient syndrome extraction rounds, leading to incorrect fixup operations.

\begin{figure}[t]
\begin{center}
\includegraphics[width=1\linewidth]{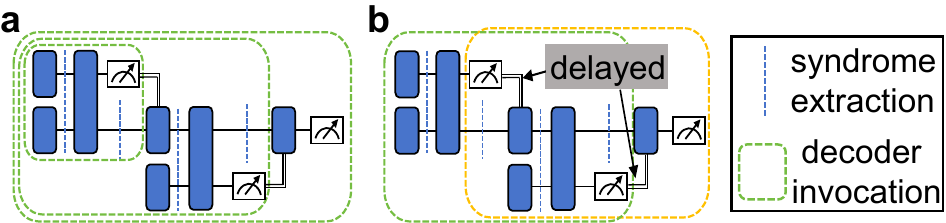}
\end{center}
\setlength{\abovecaptionskip}{0pt}
\caption{\textbf{Comparison of correlated decoding without and with window architecture in the circuit with mid-circuit measurement and feedback.} The blue blocks represent single-qubit or two-qubit logical gates and the blue dashed lines represent one round of syndrome extraction. The rectangles denote each decoder invocation and the range of syndrome data it processes.
\textbf{(a)} Without window architecture, correlated decoding is invoked after each mid-circuit measurement. 
\textbf{(b)} In sliding window architecture, correlated decoding processes syndrome data within each fixed-size window, using delayed fixup circuits to defer the decision for feedback.}
\label{fig1}
\end{figure}

Recently, a fast $T$ gate scheme called ``algorithmic fault tolerance" was proposed~\cite{zhou2024algorithmicfaulttolerancefast}, the key of which is addressing inconsistencies decoding results through classical processing. Specifically, when the measurement outcome on the magic state provided by the decoder is inconsistent with previous results, certain trivial logical Pauli operators are applied to the initial state to flip the measurement outcome on the magic state, thereby maintaining consistency. Finding these operators requires constructing and solving an additional linear system of equations during the decoding. In addition, the decoder and feedback process must be invoked after each mid-circuit measurement on magic states (see Fig.~\ref{fig1}a). Given the execution of the correlated decoding algorithm are typically slower, this increases the frequency and duration of decoding. 

More notably, this design is incompatible with window-based decoding, which stands out as a pivotal technology for decoding large volumes of syndrome data in a scalable and efficient manner~\cite{10.1063/1.1499754}. Window-based decoding divides the continuous stream of syndrome information into manageable windows, allowing for parallel processing to reduce the time complexity of decoding in large-scale quantum circuits~\cite{10.1063/1.1499754,1055873,641542,skoric2023parallel, PRXQuantum.4.040344}. In this approach, the decoder is invoked once for each fixed-size window to process the corresponding syndrome data. On the contrary, the implementation of fast $T$ gates requires invoking the decoder after each mid-circuit measurement. As window-based methods become essential for handling large-scale circuits, this inherent conflict becomes more pronounced, presenting  a significant challenge to the application of previously proposed fast $T$ gate schemes.

In this work, we design an architecture that realizes scalable constant-time logical gates for universal quantum computing, seamlessly integrating with window-based correlated decoding. This method utilizes delayed fixup circuits to postpone the fixup operation until sufficient syndrome extraction rounds have occurred, thereby ensuring fault-tolerant measurements and feedback in the non-Clifford gate implementation. While achieving constant-time logical gates and universality, the main advantage of our architecture is that it permits only a single decoder invocation for each sliding window (see Fig.~\ref{fig1}b). In other words, syndrome data within a window is processed in a single batch, thus enhancing the efficiency of the correlated decoder when handling large-scale circuits.

By dividing syndrome data into temporal sliding windows, the correlated decoding problem can be efficiently and scalably addressed for deep quantum circuits with sequentially implemented logical gates.  Nevertheless, for another widely used circuit structure in large-scale quantum computing, time-optimal quantum computation (TOQC)~\cite{fowler2013timeoptimalquantumcomputation, Gidney2021howtofactorbit, Litinski2019gameofsurfacecodes}, the correlated decoding encounters new challenges. In TOQC, logical gates are executed in parallel to the extent possible, thereby minimizing the execution time of the circuit. The trade-off of this approach is that the time overhead is converted into space overhead, increasing the number of parallel qubits. Since the time for exact correlated decoding grows exponentially with the number of logical qubits~\cite{PhysRevLett.133.240602,1055873,641542}, the decoding scale of each temporal window can easily become uncontrollable. Interestingly, we found that further dividing the windows spatially can effectively addresses this challenge. While dividing into spatial windows is not always feasible for circuits with general structures, it is particularly well-suited for TOQC. Spatial window decoding reduces the size of windows and harnesses parallelism to accelerate the decoding, thereby broadens the range of applications for window-based correlated decoding.

Adapting correlated decoding to window-based approach enables its application to large-scale quantum computation in a scalable and efficient manner. Using Shor's algorithm as an example, we explore the application of our architecture. Considering connectivity requirements, transversal gates are more feasible on platforms like ion traps~\cite{postler2022demonstration,ryan2022implementing} or neutral atoms~\cite{bluvstein2024logical}. One QEC round on ion trap platforms is approximately 100 to 1,000 times slower than on superconducting platforms~\cite{google2023suppressing,physrevx.11.041058,doi:10.1126/sciadv.1601540}, leading to a difference in clock speed of logical gates~\cite{PRXQuantum.2.010103}. For instance, factoring a 2048-bit integer on superconducting systems takes 8 hours~\cite{Gidney2021howtofactorbit}. One might infer that ion trap systems could require an impractical time of 800 to 8,000 hours. However, our analysis reveals that the overall run-time of the algorithm in ion-trap systems, when using constant-time transversal gates and TOQC circuits, is only around an order of magnitude longer than that in superconducting systems, while the space overhead does not increase significantly. Overall, we provide optimistic support for performing large-scale quantum computation tasks with acceptable space-time overhead on physical systems such as ion traps.

The remainder of this paper is structured as follows. Sec.~\ref{Preliminaries} introduce the correlated decoding and constant-time logical gate as a foundation. Sec.~\ref{delayed fixup} explains the proposed architecture for implementing scalable constant-time logical gates using delayed fixup circuits. Sec.~\ref{Correlated Decoding with Sliding Windows} delves into the details of temporal sliding window-based correlated decoding, explaining its mechanisms and advantages. In Sec.~\ref{Spatial Parallel Windows}, the concept of window decoding is further extended to spatial windows, with a focus on its distinctive suitability for TOQC. Sec.~\ref{Numerical Results} presents numerical evaluations of the performance of temporal and spatial window-based correlated decoding through two illustrative examples. Sec.~\ref{Large-Scale Quantum Computation} provides a brief analysis of the impact of these results on large-scale quantum computing, using Shor's algorithm as an example. Finally, Sec.~\ref{Discussion} concludes and discusses this work.

\section{Preliminaries} \label{Preliminaries}
This section reviews the relevant aspects of correlated decoding and explains how constant-time logical gates can be implemented within the framework of correlated decoding.

\subsection{Correlated decoding}
For CSS codes~\cite{548464}, the logical space is determined by stabilizer generators, denoted as $\{s_x\}$ and $\{s_z\}$, which represent the sets of $X$-type and $Z$-type stabilizer generators, respectively. When running multiple rounds of syndrome extraction circuits, the error syndrome $\sigma_{n,k}(s_{x/z}) \in \{+1, -1\}$ is defined as the measurement outcome of the stabilizer generator $s_{x/z}$ on the $n$-th logical qubit in the $k$-th round.

Next, we assume that the logical operations in the circuit are transversal Clifford gates, followed by several rounds of syndrome extraction or transversal measurements. In the correlated decoding, the syndrome checks (or checks in brief) $C_{n,k}(s_{x/z})$, are assigned corresponding values $c_{n,k}(s_{x/z}) \in \{+1, -1\}$ defined as the product of $\sigma_{n,k}(s_{x/z})$ and its previous result (if one exists), obtained by back-propagating the stabilizer operator through the circuit to the point where it was previously measured. Specifically, between the syndrome extraction in the $(k-1)$-th and $k$-th rounds, if a logical Pauli operation (or an idle operation) is applied, then
\begin{equation}
c_{n,k}(s_{x/z}) = \sigma_{n,k}(s_{x/z}) \sigma_{n,k-1}(s_{x/z}).
\end{equation}
If logical $H_L$ gate is applied between the $(k-1)$-th and $k$-th rounds of syndrome extractions, then
\begin{equation}
c_{n,k}(s_{x/z}) = \sigma_{n,k}(s_{x/z}) \sigma_{n,k-1}(s_{z/x}).
\end{equation}
If logical CNOT gate on qubit $n_1$ (control) and $n_2$ (target) is applied between the $(k-1)$-th and $k$-th rounds of syndrome extractions, then
\begin{equation}
\begin{aligned}
    c_{n_1,k}(s_{x}) &= 
    \sigma_{n_1,k}(s_{x}) 
    \sigma_{n_1,k-1}(s_{x}) 
    \sigma_{n_2,k-1}(s_{x}), \\
    c_{n_2,k}(s_{x}) &= 
    \sigma_{n_2,k}(s_{x}) 
    \sigma_{n_2,k-1}(s_{x}), \\
    c_{n_1,k}(s_{z}) &= 
    \sigma_{n_1,k}(s_{z}) 
    \sigma_{n_1,k-1}(s_{z}),\\
    c_{n_2,k}(s_{z}) &= 
    \sigma_{n_2,k}(s_{z}) 
    \sigma_{n_2,k-1}(s_{z}) 
    \sigma_{n_1,k-1}(s_{z}).
\end{aligned}
\end{equation}

It is more intuitive to explain correlated decoding on a hypergraph. In this representation, each vertex corresponds to a check $C_{i}$. 
Here, $C_{i}$ represents the $i$-th check in a certain sorted order of $C_{n,k}(s_{x/z})$, and its value is $c_\textit{i}$. Each hyperedge $e_j$ in the graph represents an error event, with its value $E_j$ being either 0 or 1, indicating whether the event does not occur or occurs, respectively. Such an error event will flip the values of the checks associated with the vertices of that hyperedge. The set of error event values $E_j$ that cause the flipping of the check $C_{i}$ is denoted as $I(C_i)$. Furthermore, each hyperedge is assigned a weight $w_j = -\log (p_j/1-p_j)$, reflecting the error probability $p_j$ of the error event.

As a result, the correlated decoding problem is transformed into a minimum matching problem to identify $\{E_j\}$ on a weighted hypergraph, as follows~\cite{PhysRevLett.133.240602}:
\begin{equation}
\begin{aligned}
\text{maximize} \quad & \sum_{j=1}^M  w_j E_j  \\
\text{subject to} \quad & \sum_{E_j \in I(C_i)} E_j - 2K_i = \frac{1}{2}(1 - c_i), 
\\& \qquad \qquad \quad \forall i = 1, \ldots, N, \\
& K_i \in \mathbb{Z}_{\geq 0}, \quad \forall i = 1, \ldots, N, \\
& E_j \in \{0, 1\}, \quad \forall j = 1, \ldots, M,
\end{aligned}
\end{equation}
where $M$ and $N$ represent the number of error events and checks, respectively, and $K_i$ is an integer slack variable.

According to Ref.~\cite{PhysRevLett.133.240602}, the solution to this problem can be exactly obtained using a mixed-integer programming algorithm, known as the most-likely error (MLE) decoder, which requires exponential time. Alternatively, it can be approximately solved using the hypergraph union-find (HUF) decoder~\cite{Delfosse2021almostlineartime,9682738}, which operates in polynomial time.

\subsection{Constant-time logical gate}
In the conventional FTQC framework, the decoding of each logical block is performed independently. Before executing transversal CNOT gates between logical blocks, the decoding process must first be conducted to ensure that errors are corrected before propagating through the transversal CNOT gate. Due to the imperfect measurements, this decoding requires at least $\Theta(d)$ rounds of syndrome data, where $d$ is the code distance. As a result, the time overhead for logical gates includes a factor of $\Theta(d)$ rather than being constant. It is worth mentioning that, despite differences in mechanisms, this conclusion also applies to logical operations based on lattice surgery~\cite{Horsman_2012,Litinski2018latticesurgery,Litinski2019gameofsurfacecodes}.

However, when correlated decoding is applied to the syndrome data across multiple logical blocks, the $\Theta(d)$ rounds of syndrome extraction typically required for transversal gates are no longer necessary. In correlated decoding, transversal gates and syndrome extraction are considered as a unified process, where transversal gates are viewed as automorphic mappings on the stabilizer group of the code. By tracking these mappings, syndrome information can still be effectively extracted and used for decoding. This allows decoding of a circuit consisting of a sequence of transversal gates, where on average, only one or fewer rounds of syndrome extraction are performed per layer of transversal gates, while maintaining fault tolerance~\cite{PhysRevLett.133.240602}.

Correlated decoding enables transversal gates to be implemented fast, with a time overhead of only $\Theta(1)$. For simplicity, this paper assumes that a single round of syndrome extraction is performed immediately after each transversal gate and focuses on logical qubits encoded on surface codes, although this is not strictly necessary. A set of transversal gates for surface codes includes $\{H, S, \text{CNOT}\}$. Therefore, these logical gates can be implemented in constant time. However, this set alone is insufficient for computational universality.

\begin{figure}[tbp]
\begin{center}
\includegraphics[width=1\linewidth]{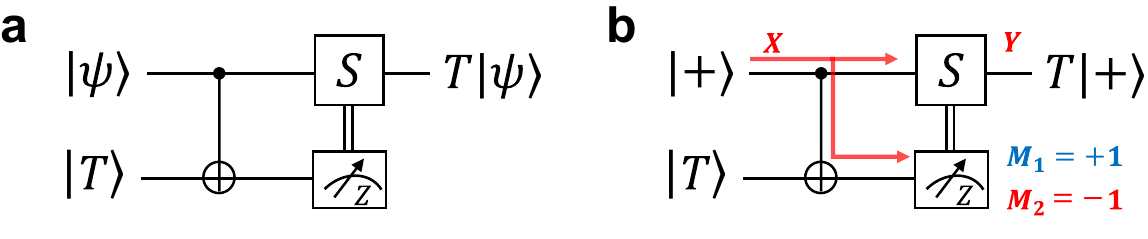}
\end{center}
\setlength{\abovecaptionskip}{0pt}
\caption{\textbf{(a)} Gadget for implementing $T$ gate consuming a magic state $\ket{T}=(\ket{0}+e^{i\frac{\pi}{4}}\ket{1})/{\sqrt{2}}$. The \( S \) gate in the circuit is classically controlled by the measurement outcome of the magic state.
\textbf{(b)} Using a trivial Pauli operator to reinterpret the measurement outcome. When twice decoding yields inconsistent results ($M_1$ and $M_2$), the measurement outcome can be reinterpreted by applying a trivial Pauli $X$ operator. The $X$ operator propagates through a CNOT gate, flipping the measurement outcome of the magic state, while leaving a $Y$ operator at the end of the circuit.}
\label{fig2.1}
\end{figure}

To implement constant-time logical gates in a universal gate set, it is beneficial to consider the $T$ gate implementation using the gadget shown in Fig.~\ref{fig2.1}a. Note that the fixup $S$ operation here is classically controlled by the measurement outcome on the magic state $\ket{T}$, which depends on decoding the syndrome data prior to the mid-circuit measurement. The challenge is that if the $T$-gadget is executed with only one round of syndrome extraction per transversal gate, the syndrome data may be insufficient to ensure fault-tolerant decoding. A naive solution is to accumulate $\Theta(d)$ rounds of syndrome data by repeatedly extracting syndromes before performing decoding and feedback on the computational qubits. However, this approach potentially makes the implementation time of the $T$ gate proportional to the code distance $d$ in the worst case, rather than constant.

Ref.~\cite{zhou2024algorithmicfaulttolerancefast} proposes a constant-time $T$ gate scheme, illustrated with the example shown in Fig.\ref{fig2.1}b. Maintaining one round of syndrome extraction per transversal gate, if the outcome of the mid-circuit measurement is later proven incorrect by subsequent decoding, a trivial $X_L$ operator is applied to the first logical qubit. This operator flips the measurement outcome and leaves only a Pauli $Y_L$ operator on the first logical qubit. This ensures that the earlier decoding inaccuracies can be effectively corrected in subsequent steps. For more general circuits with input stabilizer states, finding such logical Pauli operators requires solving a system of linear equations. Ref.~\cite{zhou2024algorithmicfaulttolerancefast} demonstrates that the probability of such a system being unsolvable decreases exponentially with increasing code distance, thereby ensuring fault tolerance. 

Remarkably, in this scheme, the decoding must be performed immediately after each mid-circuit measurement, which makes it challenging to process large amounts of syndrome data using a manageable window-based approach. First, due to the uncertainty in mid-circuit measurement positions, a fixed-step sliding window no longer adapts effectively. Secondly, if a decoding window of size $\Theta(d)$ is used for each mid-circuit measurement, these windows will have significant overlapping regions. As a result, the same syndrome data will be processed repeatedly in multiple windows, reducing decoding efficiency.

\section{Constant-time non-Clifford gates with delayed fixup circuit}\label{delayed fixup}
Here we propose an scalable constant-time implementation of non-Clifford gates using a delayed fixup circuit. We focus on surface codes to explain this approach, though it applies to other QEC codes that promise transversal CNOT gate and logical $X$ and $Z$ measurements~\cite{PhysRevLett.77.793,PhysRevA.54.1098}. 

\begin{figure}[t]
\begin{center}
\includegraphics[width=1\linewidth]{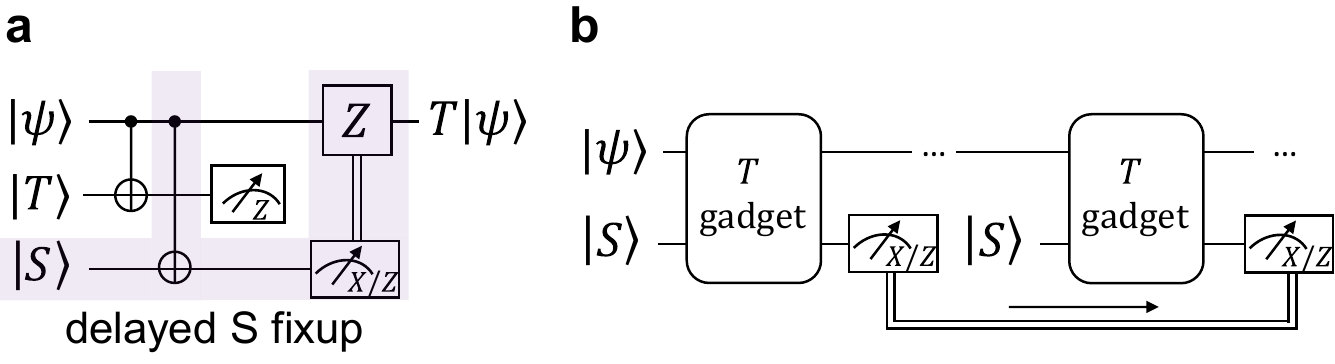}
\end{center}
\setlength{\abovecaptionskip}{0pt}
\caption{\textbf{(a)} $T$-gadget with delayed $S$ fixup. The decision for $S$ fixup is delayed until the choice of the $X$ or $Z$ measurement basis.
\textbf{(b)} When multiple non-commuting $T$ gates are applied sequentially, the measurement basis for the $\ket{S}$ state must be determined sequentially. The $T$-gadget is abstracted to represent, showing only the ancilla $\ket{S}$ state while omitting the $\ket{T}$ state.}
\label{fig3.1}
\end{figure}

When implementing non-Clifford gates such as $T$ gates or Toffoli gates on surface-code logical qubits, it is typically required to perform fixup operations classically controlled by measurements on magic states. By selecting the ancilla measurement basis, fixup decisions can be delayed. For example, in Fig.~\ref{fig3.1}a, the delayed fixup circuit uses two ancilla states, $\ket{T}=(\ket{0}+e^{i\frac{\pi}{4}}\ket{1})/{\sqrt{2}}$ and $\ket{S}=(\ket{0}+i\ket{1})/{\sqrt{2}}$, for implementing the $T$ gate and the $S$ fixup operation, respectively. If a $Z$ measurement is performed on the $\ket{S}$ state, the $S$ fixup operation is applied (up to a Pauli $Z$ operation). Conversely, if the $S$ fixup operation is not needed, an $X$ measurement is performed on the $\ket{S}$ state. Since the timing of the measurement can be arbitrarily delayed, the decision to apply the fixup operation can also be postponed until it is actually required. For circuits with multiple non-commuting $T$ gates, the $\ket{S}$ measurement basis must be chosen sequentially, as illustrated in Fig.~\ref{fig3.1}b. 

Note that delayed fixup circuits are not introduced for the first time in this work. They have been elegantly designed in several previous literatures~\cite{fowler2013timeoptimalquantumcomputation,Litinski2019gameofsurfacecodes,gidney2019flexiblelayoutsurfacecode} and have been applied in various FTQC-related works~\cite{fowler2013timeoptimalquantumcomputation,Gidney2021howtofactorbit}. However, their primary purpose in these contexts has been to enable gate parallelism for reducing time overhead, rather than to realize constant-time non-Clifford gates. These results allows us to easily utilize well-designed delayed fixup circuits to implement common non-Clifford gates in constant time.

To achieve constant-time logical gates, single rounds of syndrome extraction are performed after each transversal CNOT gate on both the computational qubits and the ancilla state $\ket{T}$. On the ancilla state $\ket{S}$, the measurement is delayed until sufficient syndrome extraction rounds have been completed. Once correlated decoding provides a reliable measurement outcome for $\ket{T}$, the basis of $\ket{S}$ measurement can be chosen (as detailed in the next section). Overall, on the computational qubits, transversal CNOT gates consume constant time overhead and $Z$ fixup can be achieved in the Pauli frame~\cite{knill2005quantum,Chamberland2018faulttolerant}, which ensures that the implementation time of the logical $T$ gate is irrelevant to $d$. 

A potential concern is that keeping the $\ket{S}$ state longer in the circuit might accumulate logical errors or increase time of qubit occupancy. However, as we will explain in the window-based decoding, with an appropriate window length, the $\ket{S}$ state persists for at most $\Theta(d)$ rounds, preventing significant increase in logical error rates or resource overhead.

In summary, the delayed fixup circuit ensures sufficient syndrome extraction rounds before fixup operations, making the decision on fixup fault-tolerant. More importantly, this architecture is compatible with window-based decoding methods (see the next section), thus addressing the scalability issue. Combined with constant-time transversal Clifford gates on surface codes~\cite{PhysRevLett.133.240602}, it is easy to confirm that this architecture achieves constant-time logical gates in a universal gate set. In addition, although here we focuses on the implementation of constant-time logical $T$ gates, generalizing the approach is straightforward. By utilizing the delayed fixup circuits proposed in Ref.\cite{gidney2019flexiblelayoutsurfacecode} and Ref.\cite{fowler2013timeoptimalquantumcomputation}, logical Toffoli gates and other non-Clifford gates can also be realized within constant time using this architecture.

\section{Correlated Decoding with Sliding Windows}\label{Correlated Decoding with Sliding Windows}
Now we introduce the temporal sliding windows in correlated decoding and explain its unique adaptability for the design in the previous section. To be precise, the window refers to a subregion composed of syndrome checks, which are arranged in temporal and spatial order within a three-dimensional space-time diagram. Hereafter, assuming the output of correlated decoding is $\{E_j\}$, we refer to the hyperedges $e_j$ satisfying $E_j = 1$ as the decoding results for each windows. 

\begin{figure}[t]
\begin{center}
\includegraphics[width=1\linewidth]{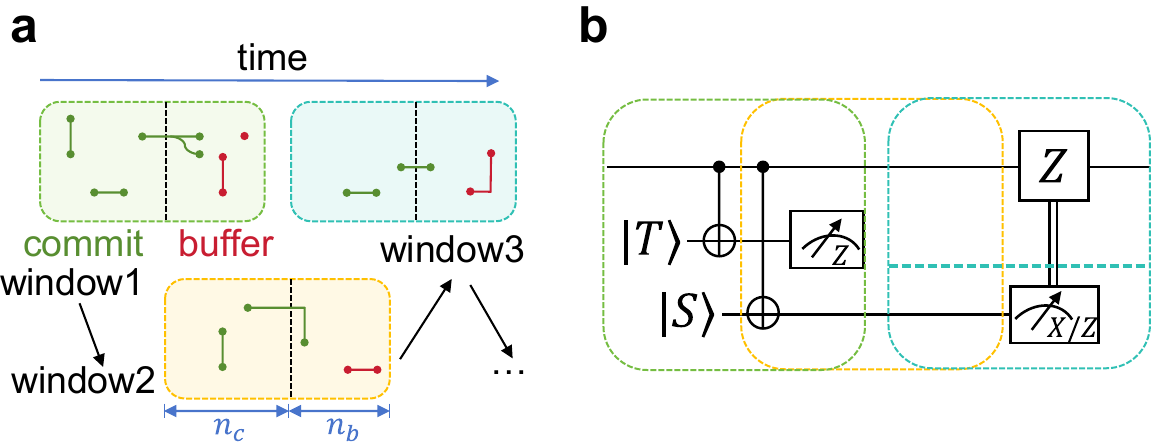}
\end{center}
\setlength{\abovecaptionskip}{0pt}
\caption{\textbf{Sliding window architecture and application in $T$-gadget.}
\textbf{(a)} Temporally feed-forward sliding window. The window slides from left to right along the temporal direction, with the spatial dimensions compressed into the other direction. The windows are processed sequentially, with each step removing the green hyperedges as reliable decoding results, while leaving the checks on the red hyperedges to be handled by the next window.
\textbf{(b)} Application of the temporal window decoding in $T$-gadget. When the measurement of $\ket{T}$ is located in the commit region of the second window, the decoder provides a reliable measurement outcome for $\ket{T}$. Based on this outcome, the measurement basis is chosen, and the measurement of $\ket{S}$ is subsequently performed. In the third window, $\ket{S}$ is decoded independently. The $\ket{S}$ state persists for at most one window length of syndrome extraction rounds.}
\label{fig4.1}
\end{figure}

In a feed-forward sliding window architecture, the window with length $n_w$ slides forward in time (see Fig.~\ref{fig4.1}a). At each step, correlated decoding is performed on $n_w$ rounds of syndrome data. The decoding result of each window is divided into two regions: a commit region and a buffer region with lengths $n_c$ and $n_b$, respectively, where $n_w=n_c+n_b$. After that, the window slides forward by $n_c$ rounds. In other words, the buffer region is the overlap between two adjacent windows. To ensure fault tolerance of the decoding, $n_w$, $n_c$, and $n_b$ are typically proportional to $\Theta(d)$.

Since the syndrome checks near the window boundary lack future syndrome information, they might be matched as a hyperedge incorrectly. Thus, only the decoding results within the commit region, which is sufficiently distant from the boundary, are considered reliable. Specifically, if all vertices of a hyperedge lie entirely within the commit region or spanning two regions, the result is preserved as final and the syndrome checks on the vertices are flipped. If all vertices of a hyperedge are entirely within the buffer region, it is deferred for the next window to resolve. 

Next, let us describe in detail how correlated decoding is executed within a window. First, we need to establish the relationship between each error event $E_j$ with its error rate $p_j$ and its corresponding hyperedge $e_j$. For each error event $E_j$, a noiseless circuit with the error event $E_j$ is simulated, and the set of checks with a value of $-1$, denoted as $C(E_j)$, is recorded. However, in a window of limited size, some checks in $C(E_j)$ may fall outside the window. Thus, in correlated decoding within a specific window, we remove checks that lie outside the current window and consider the remaining checks as the vertices of hyperedge $e_j$. If none of the vertices are within the current window, the error event $E_j$ is removed from the error event set of the current window. Additionally, hyperedges spanning both the commit and buffer regions will be removed after current decoding. Therefore, for the error event $E_j$ and hyperedges $e_j$ that spans the commit and buffer regions of a window, we remove $E_j$ from the error event set in the next window, and only consider $e_j$ and $E_j$ in the current window.

After each correlated decoding within a window, we preserve the hyperedges inside the commit region and those on the boundary of the commit and buffer region as reliable results. The syndrome information corresponding to these hyperedges is then eliminated, meaning the check values at the vertices of these hyperedges are flipped. The decoding of subsequent windows is then carried out, and the final output consists of all reliable hyperedges in each decoding as the complete decoding result.

Then, we can discuss circuits containing mid-circuit measurements and feedback more concretely. Taking the $T$-gadget in Fig.~\ref{fig4.1}b as an example, when state $\ket{T}$ measurement appears in the commit region of a sliding window (the yellow one in the figure), the decoding yields a reliable measurement outcome. Following this, the state $\ket{S}$ is measured in the appropriate basis immediately. The measurement outcome of $\ket{S}$ will depend on the decoding result from the subsequent window. In particular, in the subsequent window, since the state $\ket{S}$ does not interact with other qubits via two-qubit gates, it can be decoded independently as a single logical qubit. Note that independent decoding of $\ket{S}$ states is crucial, especially when multiple non-commuting $T$ gates with $S$ fixup are performed consecutively. If the previous $\ket{S}$ state is decoded within an entire window, the next $\ket{S}$ measurement must be delayed until the following window. For a sequence of $S$ fixups, this will clearly result in an accumulation of errors.

This example also highlights an additional advantage of window-based correlated decoding.  When there are no two-qubit gate interactions between subsets of qubits within a window, the correlated decoding can be partitioned into several independent parts. This partitioning decreases the number of variables in the correlated decoding and enables further time reductions through parallelization. Essentially, this represents a specific instance of the spatial parallel window decoding, which will be discussed in the next section.

\section{Spatial Parallel Windows}\label{Spatial Parallel Windows}

\begin{figure*}[t]
\begin{center}
\includegraphics[width=1\linewidth]{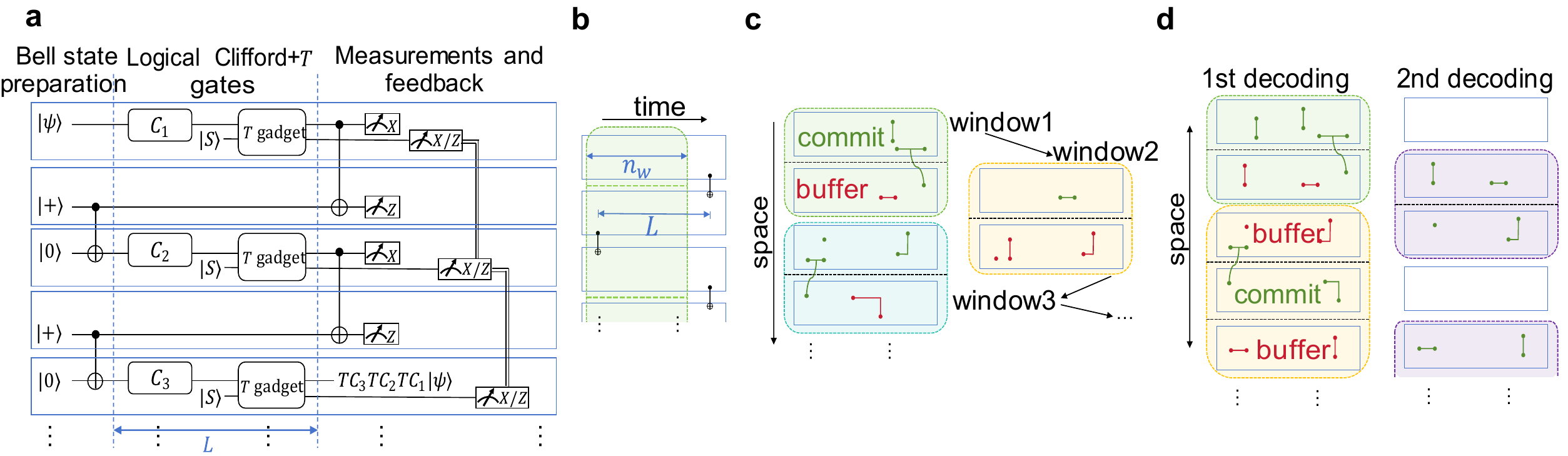}
\end{center}
\setlength{\abovecaptionskip}{0pt}
\caption{\textbf{Application of Spatial Windows in TOQC.}
\textbf{(a)} Parallel Clifford+$T$ gates in TOQC. The blue boxes divide the space into multiple blocks. The state $\ket{S}$ in the $T$ gadget and its delayed measurement are also highlighted.
\textbf{(b)} When the temporal window is shorter than the spacing between transversal CNOT gates, the window divides into independent parts naturally (by the green dot lines in the figure).
\textbf{(c)} If the temporal window covers the entire circuit, it can be spatially divided, with syndrome data processed in a feed-forward manner.
\textbf{(d)} In parallel spatial windows, decoding occurs in two stages, retaining green hyperedges as results in each.}
\label{fig5.1}
\end{figure*}

Our starting point is the concept of TOQC for large-scale quantum computing, first proposed in Ref.~\cite{fowler2013timeoptimalquantumcomputation}. Fig.~\ref{fig5.1}a illustrates an example of the TOQC circuit. Using teleportation and delayed fixup circuits, multiple groups of Clifford+$T$ gates can be executed in parallel, at the cost of requiring additional qubits. The implementation time of TOQC is primarily dominated by the depth of measurements and classical feedback, thus reducing time overhead in large-scale computing tasks. For more information on TOQC, please refer to the Appendix~\ref{toqc}.

Although TOQC reduces the time overhead, it does not lower the overall space-time cost. For instance, executing $\mathcal{O}(k)$ groups of logical gates in parallel increases the number of qubits by $\mathcal{O}(k)$ times. This naturally leads to challenges in correlated decoding of TOQC circuits. As discussed before, for circuits with large depth, correlated decoding can avoid processing too much syndrome data at once by using temporal sliding windows. However, in TOQC circuits, the circuit depth is converted into a larger number of qubits. Given that the complexity of exact correlated decoding increases exponentially with the number of qubits, directly applying correlated decoding to large TOQC circuits presents significant limitations.

Similar to temporal windows, we find that TOQC circuits can be divided into spatial windows for correlated decoding. Notice that TOQC circuits can naturally be divided into blocks spatially, each comprising Bell state preparations, logical Clifford+$T$ gates, and Bell-basis measurements and feedback (see Fig.~\ref{fig5.1}a). The blocks are interconnected by the initial and final transversal CNOT gates.

Now let us explain spatial windows in two scenarios.  First, if the depth $L$ of logical Clifford+$T$ gates in each block satisfies $L>n_w$ (recall that $n_w$ is the length of the temporal window), the initial and final transversal CNOT gates will not be in the same window (see Fig.~\ref{fig5.1}b). Consequently, this allows the temporal window to be divided into several independent parts for parallel correlated decoding. Conversely, if $L\leq n_w$, the temporal window cannot be directly divided into independent parts (see Fig.~\ref{fig5.1}c). In this case, we further divide the windows spatially, with each window containing syndrome data from $m_w$ blocks. Similarly, each window is composed of a commit region and a buffer region, containing $m_c$ and $m_b$ blocks, respectively. The intuition for spatial windows is that syndrome information between blocks that are spatially far apart becomes nearly independent. To keep the commit region away from the window boundary, the size of the buffer region is set to satisfy $m_b L=\Theta(d)$. In this setting, an error string connecting the commit region to the window boundary will contain at least $m_b L$ hyperedges. The condition $m_b L = \Theta(d)$ ensures such errors occur with exponentially low probability as $d$ increases.

Spatial window decoding can be executed in a feed-forward manner. First, correlated decoding is applied to the syndrome data within the window, preserving hyperedges located entirely within the commit region or those spanning both the commit and buffer regions as the reliable decoding results. Then the syndrome checks on the vertices of these hyperedges are flipped. Subsequently, decoding is performed for the next window in a similar manner.

Furthermore, spatial windows can be constructed and processed in parallel to reduce decoding time (see Fig.~\ref{fig5.1}d). Each window is divided into three parts: a central commit region, flanked by buffer regions on both sides. Similarly, the buffer regions are sized to satisfy $m_b L=\Theta(d)$. Correlated decoding is performed once for each window and then the processed commit regions naturally divide the remaining areas into several independent parts. Consequently, once more decoding can be performed in parallel for each part.  This process is similar to the parallel window method in Ref.~\cite{skoric2023parallel}, except that the windows here are divided spatially.

Here, it is beneficial to discuss the size of both temporal and spatial windows. For temporal windows without TOQC implementation, their size is immediately apparent. Each temporal window processes syndrome data from $\Theta(d)$ rounds on $n$ qubits at a time. We assume that such size of correlated decoding remains acceptable. However, in TOQC circuits, a spatial window includes multiple $n$-qubit blocks. One might wonder whether the size of a spatial window still remains acceptable.

To ensure fault tolerance, the distance between the commit region and the window boundary must be sufficiently large. To be precise, the distance refers to the minimum number of hyperedges required to connect two vertices. In fact, this distance only depends on the size of the buffer region. For temporal and spatial windows, $n_b$ and $m_b$ can be set to $(d+1)/2$ and $(d+1)/2L$, respectively. This ensures that the probability of an error chain connecting the commit region to the window boundary is $\mathcal{O}(p^{(d+1)/2})$, which will not significantly contribute to the logical error rate $P_L$. On the other hand, the size of the commit region is not restricted. Typically, to output as many reliable decoding results as possible while keeping the overall window size manageable, the size of the commit region is set equal to that of the buffer region.

Now, under these settings, let us calculate the sizes of the temporal and spatial windows. Suppose the temporal window includes $n$ qubits in the circuit, then the space-time volume of the temporal window is $n(n_c + n_b) = n(d+1)$. For the spatial window, assume that each block in the TOQC circuit has $n$ qubits and a logical gate depth of $L$. The space-time volume of the spatial window is also $n(m_c + m_b)L = n(d+1)$. This result dispels concerns about the size of spatial windows, demonstrating that both spatial and temporal windows can perform correlated decoding within a limited scale.

Finally, we would like to emphasize that spatial windows does not apply to more general circuits. Our work only demonstrates its applicability specifically to the structure of TOQC circuits. This is because, in TOQC circuits, the CNOT gates between blocks are very limited and appear only at the beginning and end of the circuit, which means that the correlation of error syndromes is influenced by spatial distance.

\section{Numerical Results}\label{Numerical Results}
\begin{figure*}[tbp]
\begin{center}
\includegraphics[width=1\linewidth]{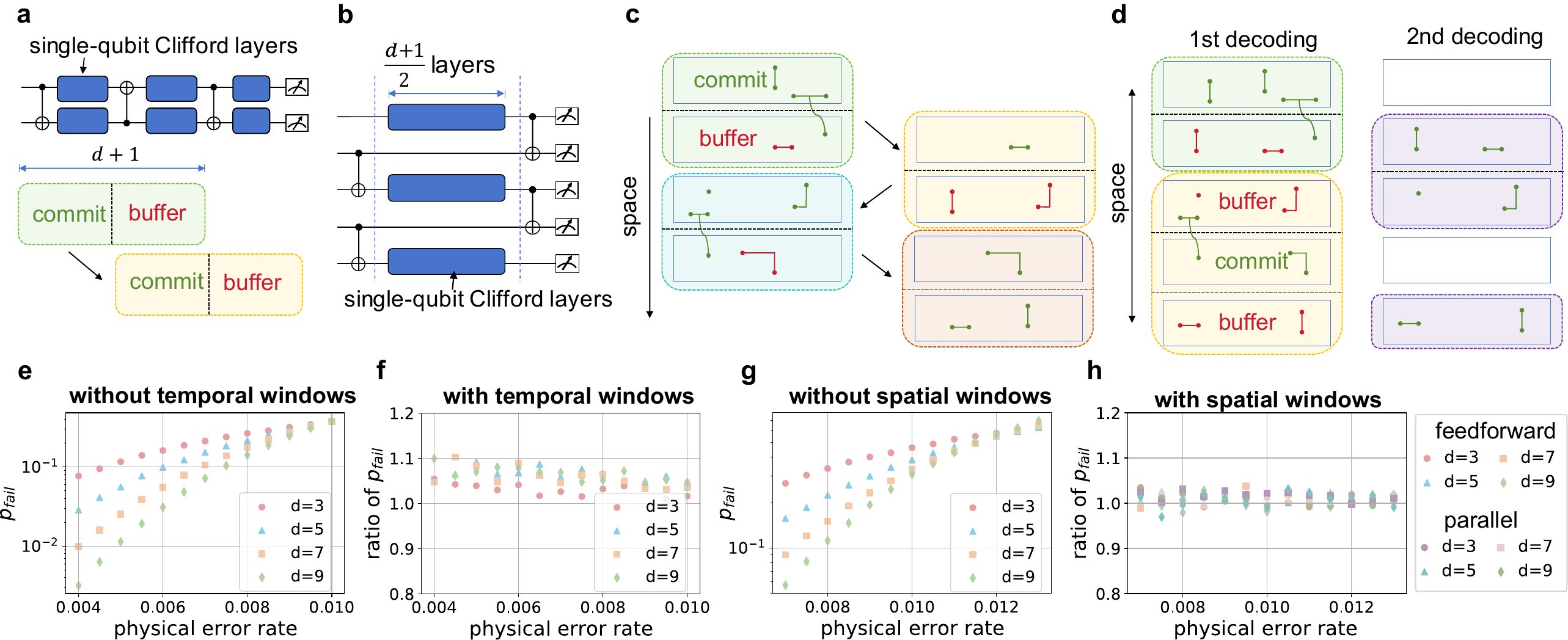}
\end{center}
\setlength{\abovecaptionskip}{0pt}
\caption{\textbf{Numerical simulation of two examples.} (a-d) show the circuits we simulated along with the corresponding window-based decoding methods. (e-h) present the results of the numerical simulations. The statistical error of one standard deviation for each data point in the figure is no greater than the size of the marker for that point.
\textbf{(a)} Circuit and temporal feed-forward sliding window for the first example. Each window contains $d+1$ rounds of syndrome data.
\textbf{(b)} Circuit for the second example, with the similar structure as the TOQC.
\textbf{(c)} Feed-forward spatial windows for decoding the second example. Each spatial window contains syndrome data on two logical qubits.
\textbf{(d)} Parallel spatial windows for decoding the second example.
\textbf{(e)}  Decoding failure rate $p_{\text{fail}}$ at different code distances as a function of the physical error rate without window-based decoding for the first example.
\textbf{(f)}  The ratio of decoding failure rates with temporal windows to those without.
\textbf{(g)}  Decoding failure rate $p_{\text{fail}}$ at different code distances as a function of the physical error rate without window-based decoding for the second example.
\textbf{(h)}  The ratio of decoding failure rates with spatial windows to those without, including both feed-forward and parallel types.}
\label{fig6}
\end{figure*}

We present two examples (see Fig.~\ref{fig6}) to illustrate the performance of the window-based correlated decoding. Here, the logical qubits in the circuit are encoded on surface codes, enabling transversal CNOT gates and single-qubit Clifford gates from the set $\{X, Y, Z, H\}$. Fig.~\ref{fig6}~(a-b) illustrate the specific circuits for both examples, where syndrome extraction is performed for single round after each logical gate. The division of windows is shown in Fig.~\ref{fig6}a and (c-d). We also provide details of the noise model and simulation methods in Appendix ~\ref{simulation}.

In the first example, the circuit consists of two logical qubits and a series of transversal CNOT gates and single-qubit Clifford layers, with a total depth of $3(d+1)/2$. The sliding window length is set to $n_w=d+1$, where the commit and buffer regions each occupy half. The entire circuit is decoded within two temporal windows sequentially. 

In the second example, the circuit consists of five logical qubits and has a structure similar to TOQC with $L=(d+1)/2$. We tested the decoding process under two window architectures, feed-forward and parallel windows. In the feed-forward window architecture, each window contains syndrome data from two blocks. Correlated decoding is performed sequentially across four windows, with the window sliding by one block in the spatial direction each time. In the parallel window architecture, decoding is carried out in two steps: first, two disjoint spatial windows are decoded, followed by independent correlated decoding in the remaining buffer regions of the two parts.

In both two examples, our setup ensures that the distance from the commit region to the window boundary is at least $\frac{d+1}{2}$. This condition ensures that the additional logical error rate introduced by the windowing method is $\mathcal{O}(p^{(d+1)/2})$, on the same order as the logical error rate of the surface code encoding itself. For comparison to window-based decoding, we also performed correlated decoding without windows, meaning a single decoding process is applied to all syndrome data from the entire circuit. We define the decoding failure rate $p_{\text{fail}}$ as the probability of a logical Pauli error occurring on any logical qubit after decoding. Fig.~\ref{fig6}e and Fig.~\ref{fig6}g show the failure rates of decoding without the window architecture for the two examples. In comparison, Fig.~\ref{fig6}f and Fig.~\ref{fig6}h display the ratio of the failure rates with and without the window architecture for the same examples. The results indicate that the performance of window-based correlated decoding is almost identical, with only a very slight increase in $p_{\text{fail}}$.

\section{Application in Large-Scale Quantum Computation}\label{Large-Scale Quantum Computation}
In this section, we aim to provide a preliminary analysis of the impact of constant-time quantum gates on large-scale quantum computation, using the Shor algorithm as an example. Our analysis builds upon Ref.~\cite{Gidney2021howtofactorbit}, comparing the time and space overhead of implementing Shor's algorithm using constant-time transversal gates in ion trap systems versus the lattice surgery~\cite{Horsman_2012,Litinski2018latticesurgery,Litinski2019gameofsurfacecodes} in superconducting systems.

Some basic assumptions include the following. First, it is assumed that the error rates for physical gates, initialization and measurements are equal for the two systems, while the ion-trap system can provide the required connectivity. Second, the QEC round time for the ion-trap system is 100 to 1,000 times that of the superconducting system~\cite{google2023suppressing,physrevx.11.041058,doi:10.1126/sciadv.1601540}. Additionally, in ion trap systems, the reaction time, defined by measurement plus classical processing and feedback time, is about an order of magnitude longer than in superconducting systems due to longer measurement time~\cite{gidney2019flexiblelayoutsurfacecode, PhysRevApplied.12.014038,PhysRevLett.126.010501,crain2019high}.

Ref.~\cite{Gidney2021howtofactorbit} identifies the two most time-consuming subroutines as the ripple-carry adder~\cite{cuccaro2004newquantumripplecarryaddition} and the lookup table~\cite{PhysRevX.8.041015}, both of which are also important subroutines in other algorithm~\cite{PRXQuantum.2.030305,PhysRevResearch.3.033055}. The ripple-carry adder, executable via TOQC circuits, is limited by the reaction time. At this point, fast transversal gates offer negligible time savings since the computation time is proportional to the measurement depth. In contrast, the implementation time of lookup table is dominated by multi-target CNOT gates, also known as Clifford-limited~\cite{gidney2019flexiblelayoutsurfacecode}. With fast transversal gates and ancilla cat states, multi-target CNOT gates can be implemented in a single QEC round, mitigating the effect of slower QEC rounds in ion trap systems by approximately a factor of code distance $d$. Given the above factors, the total time of these two subroutines is approximately an order of magnitude longer in ion trap systems (see Appendix~\ref{estimation} for a more specific calculation).

Moreover, fast transversal gates also reduce the space overhead. Since the algorithm is executed with fewer QEC rounds by fast transversal gates, the accumulated logical error rate on the computational qubits will be lower, thus relaxing the requirement for the code distance $d$. This reduction in space overhead is approximately 14\% when reducing the code distance $d$ from 27 to 25 (see Appendix~\ref{estimation}). Additionally, systems with flexible connectivity will further reduce qubit requirements by saving routing qubits and allowing flexible qubit reuse. 

However, the space occupied by magic state distillation must be considered~\cite{Gidney2019efficientmagicstate}. The differing ratios of QEC round time and reaction time between ion trap and superconducting systems lead to variations in the generation and consumption rates of magic states. In ion trap system with fast transversal gates and TOQC circuits, magic states will be consumed faster, measured in units of QEC rounds, which indicates the distillation rate needs to increase. Fortunately, fast transversal gates applied in distillation can theoretically reduce space-time overhead by $\Theta(d)$ times~\cite{zhou2024algorithmicfaulttolerancefast}. Given that  distillation protocol based on lattice surgery is highly optimized~\cite{Litinski2019magicstate}, the actual factor of reduction in space-time overhead will possibly be less than $d$. Considering these factors comprehensively, although distillation factories require more space, the overhead will remain acceptable. Overall, using fast transversal gates in ion trap systems yields space overhead of Shor's algorithm comparable to using lattice surgery in superconducting systems~\cite{Gidney2021howtofactorbit}, as corroborated by the calculations in Appendix~\ref{estimation}.

\section{Conclusion and Discussion}\label{Discussion}
In this work, we present an architecture that integrates window-based correlated decoding with constant-time logical gates for universal quantum computing. Compared to the original correlated decoding, window-based correlated decoding significantly enhances scalability and efficiency, offering broad applicability for constant-time logical gates in large-scale quantum computing scenarios.

When discussing time-based windows, we only considered feedforward windows and do not address parallel windows. However, we also note that such parallelism for temporal windows has already been achieved in previous works on window-based decoding, as seen in Refs.~\cite{skoric2023parallel, PRXQuantum.4.040344}. For window-based correlated decoding, there are no fundamental difficulties in applying this approach. Hence, adapting temporal windows for parallelism in our work remains feasible. On the contrary, as we mentioned, applying spatial windows and achieving parallelism is not always feasible for general circuit structures. Intuitively, the division of spatial windows appears to be related to the degree of transverse CNOT connectivity between the blocks of the circuit. Investigating the connection between spatial windows parallelism and the circuit structures remains an interesting question.

Additionally, preliminary analysis of Shor's algorithm in ion trap systems suggests promising potential, indicating the value of constant-time logical gates in FTQC. However, we also acknowledge that a particularly detailed and accurate analysis is beyond the scope of this work. Therfore, we leave more detailed resource evaluations and optimizations to be explored in future work.

\begin{acknowledgments}
We thank the Gurobi Optimization team for providing the Gurobi software package~\cite{gurobi}, which was used for solving optimization problems in this work. This work was supported by the National Natural Science Foundation of China (Grant No. 12034018).
\end{acknowledgments}

\appendix

\section{Time-optimal quantum computation (TOQC) circuit}\label{toqc}
\begin{figure}[t]
\begin{center}
\includegraphics[width=1\linewidth]{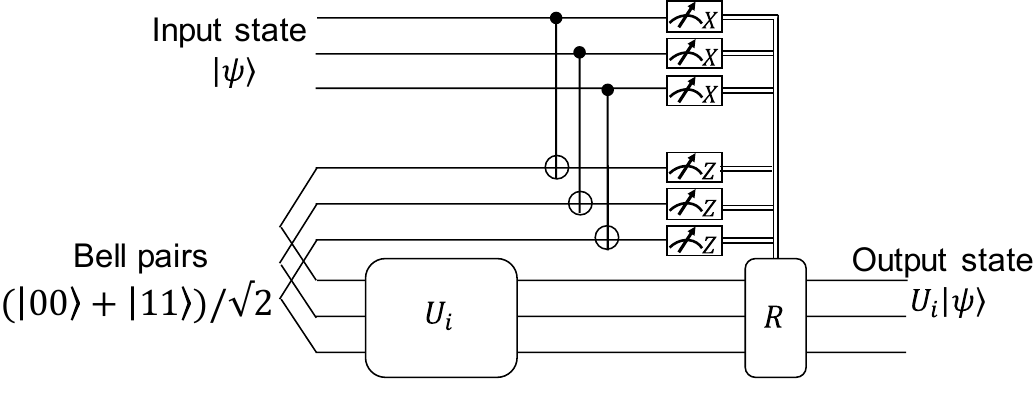}
\end{center}
\setlength{\abovecaptionskip}{0pt}
\caption{\textbf{Teleportation circuit in TOQC.}}
\label{figs1}
\end{figure}

Here, we explain additional details of the TOQC circuit mentioned in the main text. Generally, a TOQC circuit consists of two components,  the teleportation circuit (see Fig.~\ref{figs1}) and the delayed choice circuit. When the input state $\psi$ is an $n$-qubit state, the operation $U_i$ is applied to a block composed of $n$ Bell pairs. Afterward, Bell basis measurements are performed Based on the theory of teleportation, it can be easily verified that $U_i$ is successfully applied to the state $\psi$, up to a recovery operation $R$. When $U_i$ is a Clifford operation, the recovery operator $R$ is an element of the $n$-qubit Pauli group.

When $U_i$ is a Clifford operation combined with a $T$-gadget, delayed $S$ fixup can be implement to ensure that the recovery operator $R$ remains a Pauli operator and can be tracked in the Pauli frame~\cite{knill2005quantum,Chamberland2018faulttolerant}. The recovery operator $R$ will affect subsequent recovery operators and the measurement basis of the $S$ fixups. For example, if the recovery operator $X$ is generated, the equation $XTX = T^\dagger = TSZ$ implies that the $S$ gate and $Z$ gate must be applied afterward, which will be reflected in the measurement basis for future $S$ fixups and in the Pauli frame. Therefore, at the end of the circuit, the measurement basis of the $S$ state will be determined sequentially.

In the Shor's algorithm example mentioned in the main text, the non-Clifford gates in the TOQC circuit are typically Toffoli gates. The parallel execution of Toffoli gates can be achieved using delayed CZ gates with the similar methods in the TOQC, as discussed in Appendix \ref{lookup}.

Additionally, note that the TOQC circuit discussed in the main text is not entirely identical to the circuits in Ref.~\cite{fowler2013timeoptimalquantumcomputation}, which Ref.~\cite{fowler2013timeoptimalquantumcomputation} can be regarded as a more general form of TOQC. In the general form, spatial window decoding is also applicable. This is because the general TOQC circuit shares similar structure with the one discussed in the main text. Specifically, the TOQC circuit can be divided into spatial blocks, and transversal CNOT gates between blocks only appear at the beginning or end of the circuit. As a result, the discussion in the main text is also valid for the TOQC described in Ref.~\cite{fowler2013timeoptimalquantumcomputation}.

\section{Details of numerical simulations}\label{simulation}
Numerical simulations are conducted on two examples in the main text. We simulate the circuits under circuit-level noise model to obtained syndrome data. Specifically, the depolarizing Pauli noise channels are defined as follows:
\begin{equation}
\begin{aligned}
\mathcal{E}_{1}(\rho_1)&=(1-p)\rho_1+(p/3)\sum_{P\in\{X,Y,Z\}}P\rho_1P,\\
\mathcal{E}_{2}(\rho_2)&=(1-p)\rho_2+(p/15)
\times \\ &\sum_{\substack{P_1,P_2\in\{I,X,Y,Z\},\\P_1\otimes P_2\neq I \otimes I}}P_1\otimes P_2\rho_2P_1\otimes P_2,
\end{aligned}
\end{equation}
where $\rho_1$ and $\rho_2$ are single-qubit and two-qubit density matrices respectively. In the simulated circuits, we apply $\mathcal{E}_{1}$ after single-qubit gates, and idle operations, and $\mathcal{E}_{2}$ after two-qubit gates. Additionally, the measurement outcome and state initialization flips with the probability of $p$. We simulate the circuits in the Heisenberg representation~\cite{gottesman1998heisenberg}, randomly inserting Pauli noise with corresponding probability. These Clifford circuits can be efficiently simulated in polynomial time, according to the Gottesman-Knill theorem~\cite{nielsen2010quantum,PhysRevA.70.052328}. 

Then, correlated decoding is applied within each window. The decoding process uses an MLE decoder to obtain the exact solution for correlated decoding, implemented through the \textit{Gurobi} solver~\cite{gurobi}. One difference from the original correlated decoding literature is that we pre-merge several error events $\{E_i\}$ corresponding to the same hyperedge and assign an error probability based on the likelihood of an odd number of these events occurring. This approach accounts for the equivalence of error events. We expect the results after this pre-merging process to be closer to the exact solution for the decoding problem, i.e., most-likelihood decoding~\cite{PhysRevA.90.032326}. Additionally, the reduction in variables within the set $\{E_i\}$ implies a lower decoding time complexity.

For each point of Fig.~\ref{fig6}~e-g in the main text, at least 10,000 decoding failure samples are collected. As a result, the standard deviation of $p_{\text{fail}}$ is less than 0.01$p_{\text{fail}}$. In the figure of main text, the standard deviation for each data point are smaller than the sizes marked in the figure, hence the error bar are not displayed.

\section{Implementation of ripple-carry adder and lookup table}\label{lookup}
\begin{figure}[t]
\begin{center}
\includegraphics[width=1\linewidth]{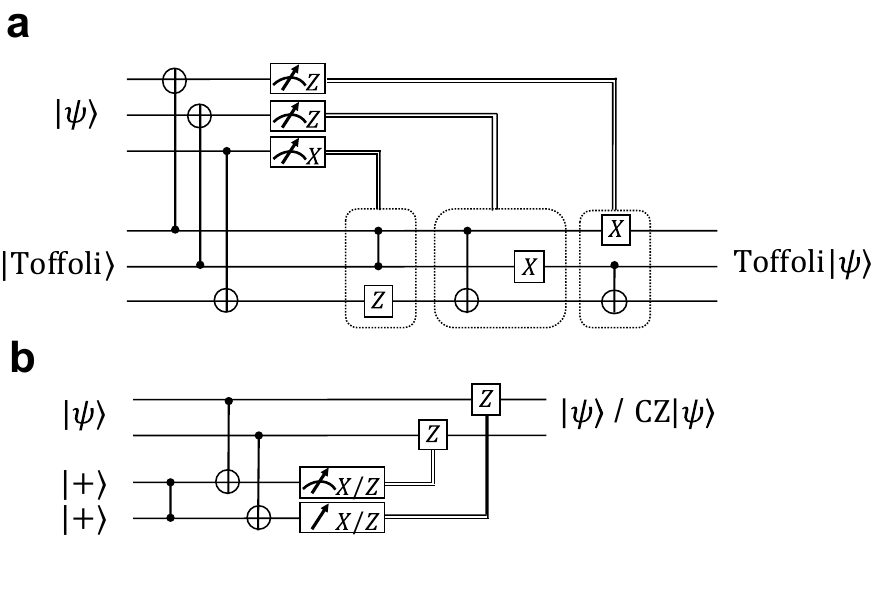}
\end{center}
\setlength{\abovecaptionskip}{0pt}
\caption{\textbf{(a)} Gadget circuit for implement Toffoli gate from \cite{PhysRevA.87.032321} with $\ket{\text{Toffoli}}=\text{Toffoli}\ket{+}\ket{+}\ket{0}$.
\textbf{(b)} Delayed-choice CZ circuit from \cite{gidney2019flexiblelayoutsurfacecode}.}
\label{figs3}
\end{figure}

In the main text, we discussed the two most time-consuming subroutines in Shor's algorithm: the ripple-carry adder~\cite{cuccaro2004newquantumripplecarryaddition} and the lookup table (or QROM)~\cite{PhysRevX.8.041015}. Their specific circuit implementations can be found in Ref.~\cite{gidney2019flexiblelayoutsurfacecode} and Ref.~\cite{haner2022spacetimeoptimizedtablelookup}. The non-Clifford gate in the circuits of these two subroutines is the Toffoli gate. For surface codes and other quantum error correction codes with transversal CNOT gates, the Toffoli gate can be implemented using the gadget circuit in Fig.~\ref{figs3}a, where the fixups are CZ and CNOT gates. To construct a TOQC circuit containing Toffoli gates, one can use the delayed CZ correction circuit shown in Fig.~\ref{figs3}b, or equivalently implement a CCZ gate using the auto-CZZ ancilla state from Ref.~\cite{gidney2019flexiblelayoutsurfacecode}.

\begin{figure}[tbp]
\begin{center}
\includegraphics[width=1\linewidth]{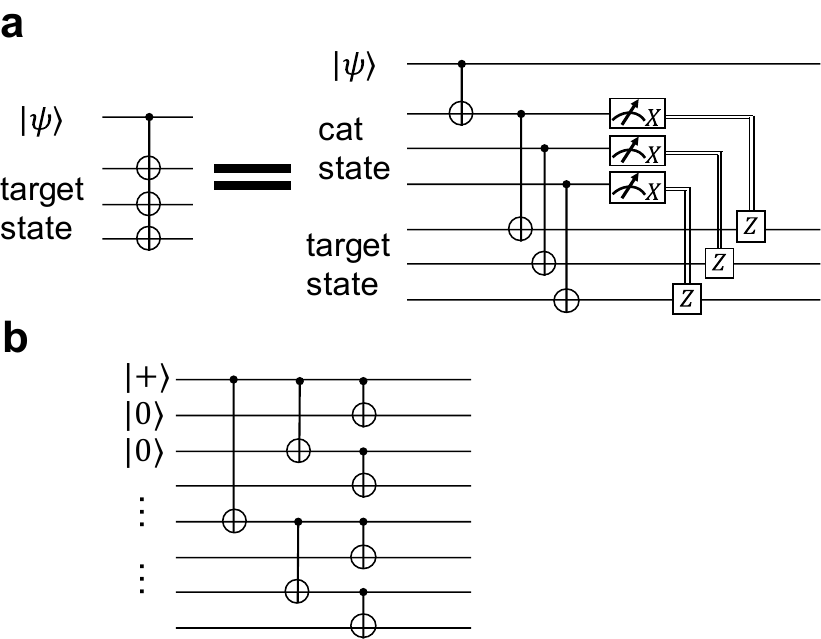}
\end{center}
\setlength{\abovecaptionskip}{0pt}
\caption{\textbf{(a)} Implementation of multi-target CNOT gate with ancilla cat state.
\textbf{(b)} Circuit for preparing $n$-qubit cat state in $\lceil \log n \rceil$ layers.}
\label{figs4}
\end{figure}

For the lookup table, another complexity lies in the implementation of multi-target CNOT gates. Without connectivity constraints, multi-target CNOT gates can be implemented using ancilla cat state $(\ket{0}^{\otimes n}+\ket{1}^{\otimes n})/\sqrt{2^n}$ followed by two layers of CNOT gates (only one layer on the computaional qubit), as shown in Fig.~\ref{figs4}a. Additionally, the $n$-qubit cat state can be prepared in $\lceil \log n \rceil$ layers of CNOT gates, as shown in Fig.~\ref{figs4}b. Here we assume that the platform allows for all-to-all connectivity. Note that in solid-state quantum systems, such as superconducting system, the time complexity of preparing an $n$-qubit cat state is $O(n)$, assuming only nearest-neighbor physical CNOT gates are allowed. Furthermore, multi-target CNOT gates cannot be heavily parallel, as routing qubits will be occupied.

\section{A More Specific Estimation of the Space-Time Overhead for Shor's Algorithm}\label{estimation}
Our estimate follows the framework laid out in Ref.~\cite{Gidney2021howtofactorbit}, which elaborates on the execution process of the Shor's algorithm within the lattice surgery framework of superconducting systems. Actually, we did not conduct a comprehensive analysis of all detailed steps of Shor's algorithm, but instead focused on the most time-consuming part, the lookup addition, which accounts for the vast majority of the space-time overhead when factoring large integers. 

Notably, our evaluation is not a complete reconstruction but rather focuses on comparing with the lattice surgery method, examining how constant-time transversal gates impact the overhead in various aspects. At the algorithm level, the analysis is the same as Ref.~\cite{Gidney2021howtofactorbit}. In other words, we maintain the same process flow of the algorithm as Ref.~\cite{Gidney2021howtofactorbit} but employ constant-time transversal gates rather than lattice surgery for the circuit implementation. In Ref.~\cite{Gidney2021howtofactorbit}, several optimizations are made for the practical execution of Shor's algorithm to factor a 2048-bit integer. Note that the parameters in this process are selected based on lattice surgery techniques for only superconducting systems. Thus, our analysis serves as a rough reference, leaving room for potential further optimization.

First, we clarify several important assumptions. It is assumed that ion trap systems allow arbitrary connectivity while maintaining the same error rate as superconducting systems. Besides, the time required for one round of QEC in ion trap systems is assumed to be $\gamma$ times longer than that of superconducting systems. Based on existing experimental data and theoretical projections, a $\gamma$ value between $10^2$ and $10^3$ is a reasonable assumption~\cite{google2023suppressing,physrevx.11.041058,doi:10.1126/sciadv.1601540}. Moreover, the reaction time in ion trap systems is estimated to be $\eta$ times that of superconducting systems. In superconducting systems, the reaction time is mainly constrained by the time needed for classical processing and feedback, generally estimated to be around 10 $\mu$s~\cite{gidney2019flexiblelayoutsurfacecode,Gidney2021howtofactorbit}. For ion trap systems, the reaction time encompasses both the measurement and classical processing, and feedback, estimated to range between 20 and 200 $\mu$s~\cite{PhysRevApplied.12.014038,PhysRevLett.126.010501,crain2019high}. Therefore, $\eta$ is assumed to fall between 2 and 20. For simplicity, we assume that $\gamma$ and $\eta$ scale proportionally, i.e., $\gamma/\eta = 50$. Additionally, the physical error rate is assumed to be $10^{-3}$, and the code distance for the computational qubits is set at $d = 27$, which, as shown in Ref.~\cite{Gidney2021howtofactorbit}, can provide a sufficiently low logical error rate for factoring 2048-bit integers.

Now, we can address the time overhead of Shor's algorithm. The time complexity is mainly driven by numerous lookup addition operations, including both the lookup phase and the addition phase. According to Ref.~\cite{Gidney2021howtofactorbit} and Ref.~\cite{gidney2019flexiblelayoutsurfacecode}, the time for the lookup phase is roughly proportional to the time taken for multi-target CNOT gates, estimated at 14 ms in superconducting systems. As discussed earlier, with fast transversal gates, multi-target CNOT gates can be executed in a single QEC round (rather than $d$ rounds) on the computational qubit. Thus, the lookup phase time in ion trap systems is estimated to be $14\text{ms} \times \gamma/d =$ 51.9 to 519 ms. In the addition phase, since the circuit is implemented using TOQC, the reduction in time overhead from fast transversal gates is almost negligible. Therefore, the addition phase time in ion trap systems is estimated to be $22\text{ms} \times \eta =$ 44 to 440 ms, where 22 ms is the corresponding time for superconducting systems.

Summing these two phases, the total time for the lookup addition operation in ion trap systems is 2.7 to 27 times longer than in superconducting systems. This ratio can serve as a rough estimate of the overall time difference between the two systems when factoring a large integer using Shor's algorithm, with ion trap systems requiring approximately one order of magnitude more time.

Regarding space overhead, we mainly focus on optimizing the computational qubits and the preparation of auto-CCZ states. In other words, we assume that the space overhead for other parts (such as registers in the lookup table) remains comparable. This assumption overlooks the differences in routing qubits between the two systems, making it more conservative for ion trap systems.

First, according to the setup in Ref.~\cite{Gidney2021howtofactorbit}, factoring a 2048-bit integer requires 6,200 computational logical qubits. As mentioned in the main text, the advantage of fast transversal gates lies in their ability to reduce the error rate. For surface codes, the logical error rate per QEC round is estimated as $P_L(d)\propto (p/p_{th})^{(d+1)/2}$~\cite{PhysRevA.71.022316}, where we assume the physical error rate $p$ and the threshold $p_{th}$ are $10^{-3}$ and $10^{-2}$, respectively. We have calculated that, in ion trap systems, a lookup addition takes 95.9-959 ms, which includes 959 QEC rounds. In contrast, this value is $36 \text{ms}/1 \mu\text{s}$= 36,000 rounds in superconducting systems. This implies a reduction in the cumulative logical error rate of computational qubits by approximately 37 times. Consequently, the code distance can be reduced from 27 to 25, resulting in about a 14\% reduction in space overhead, while the overall error rate remains no higher than that in superconducting systems. One might worry that Toffoli gates are excluded from this discussion, as their error rate is also affected by the fidelity of CCZ states. However, note that when factoring large integers using Shor's algorithm, CCZ states typically require two levels of distillation~\cite{Gidney2021howtofactorbit}. After two levels of distillation, the fidelity of the output CCZ states is dominated by the Clifford operations, rather than the fidelity of the input states in the distillation protocol~\cite{Gidney2019efficientmagicstate,Litinski2019magicstate}. Therefore, by implementing these Clifford operations with fast transversal gates, the equation of error rate still applies to Toffoli gates. 

Additionally, the preparation of auto-CCZ states also contributes to space overhead. Note that these qubits can be reused. Therefore, assuming the required time for this part remains constant, reducing space-time overhead directly implies a reduction in space overhead. First, high-fidelity CCZ states are obtained through two levels of distillation~\cite{Gidney2019efficientmagicstate}. In superconducting systems, this roughly requires $16\times113$ logical qubits. Compared to the original Ref.~\cite{Gidney2019efficientmagicstate}, this overhead has already been reduced by about 8 times in Ref.~\cite{Litinski2019magicstate}. With fast transversal gates, this overhead is expected to be further reduced by a factor of $\Theta(d)$~\cite{zhou2024algorithmicfaulttolerancefast}, which we conservatively estimate as 5 times. Second, the CCZ states need to be converted into auto-CCZ states, a process that involves several CZ gates~\cite{gidney2019flexiblelayoutsurfacecode}. In superconducting systems, this process requires around $6\times113$ logical qubits. Fast transversal gates can speed up these CZ gates by a factor of $d$, which is equivalent to reducing space overhead by a factor of $d$. Finally, the space overhead for preparing auto-CCZ states needs to be multiplied by $\gamma/\eta$. This is because the speed of generating and consuming auto-CCZ states is proportional to $\gamma$ and $\eta$ (at most), respectively. Thus, space overhead needs to be expanded by $\gamma/\eta$ to match the generation and consumption rates of auto-CCZ states. After all calculations, the space overhead in ion trap systems is actually lower, roughly 99.6\% of that in superconducting systems. 

Note that these results are based on several conservative assumptions. Clearly, these evaluations are quite rough and could be further refined. However, as we have pointed out, a particularly thorough analysis is beyond the scope of this work and will be left for future studies.

\bibliographystyle{apsrev4-2}
\bibliography{ref}

\end{document}